\documentclass[aps,pre,onecolumn]{revtex4}
\usepackage{graphicx, bm, bbm,amsmath, amssymb, epsfig}
\usepackage{latexsym}
\usepackage{xcolor}

\pdfoutput=1
\newcommand{\nn}{\noindent}

\begin{document}
\title{Semi-implicit methods for the dynamics of elastic sheets}
\author{Silas Alben$^{1,*}$, Alex A. Gorodetsky$^{2}$, Donghak Kim$^{2}$, and Robert D. Deegan$^{3}$}
\affiliation{$^1$Department of Mathematics, 
$^2$Department of Aerospace Engineering, 
$^3$Department of Physics \& Center for the Study of Complex Systems
\\University of Michigan,
Ann Arbor, MI 48109, USA}
\email{alben@umich.edu}


\begin{abstract}
Recent applications (e.g. active gels and self-assembly of elastic sheets) motivate the need to efficiently simulate the dynamics of thin elastic sheets. We present semi-implicit time stepping algorithms to improve the time step constraints that arise in explicit methods while avoiding much of the complexity of fully-implicit approaches. For a triangular lattice discretization with stretching and bending springs, our semi-implicit approach involves discrete Laplacian and biharmonic operators, and
is stable for all time steps in the case of overdamped dynamics. For a more general finite-difference formulation that can allow for general elastic constants, we use the analogous approach on a square grid, and
find that the largest stable time step is two to three orders of magnitude greater than for an explicit scheme. 
For a model problem with a radial traveling wave form of the reference metric, 
we find transitions from quasi-periodic to chaotic dynamics as the sheet thickness is reduced, wave amplitude is increased, and damping
constant is reduced. 
\end{abstract}




\maketitle
\section{Introduction}

In recent years there have been various studies of how spatial variations in the composition of a thin sheet can produce global conformational changes. Examples include the appearance of spontaneous curvature due to strain variations across the thickness of the sheet~\cite{calladine:1983a,freund:2004a,armon:2011a} or non-Euclidean reference metrics induced by in-plane strain variations~\cite{audoly:2003a,santangelo:2009a,gemmer:2011a,sharon:2012a,sharon:2016a}. 
A theory of incompatible elastic plates~\cite{efrati:2009a,efrati:2013a} has been developed to determine equilibrium configurations of such sheets.  Related approaches have been used to develop self-folding origami gels~\cite{na:2015a}.  A catalog of responsive materials are now available for investigating the mechanics of thin sheets, including non-uniform responsive gel sheets~\cite{klein:2007a,kim:2012a,wu:2013a}, sheets of nematic elastomers with a non-uniform director field~\cite{ware:2015a,white:2015a}, responsive gels combined with oriented micro-rods~\cite{gladman:2016a}, and sheets in confined geometries~\cite{lin:2009a,roman:2010a,aharoni:2010a,paulsen:2017a}. 

The {\it dynamics} of responsive gel sheets were the focus of work by Yoshida and collaborators, who synthesized a gel that locally swells in response to chemical waves propagating entirely within the gel ~\cite{yoshida:1996a}.  They used self-oscillating gels in various narrow strip geometries to make a variety of soft machines~\cite{maeda:2007a,tabata:2002a,tabata:2003a,maeda:2008a,shiraki:2012a}. The Balazs group used poroelastic simulations of self-oscillating gels to demonstrate additional examples of soft machines with uniaxial bending or isotropic swelling~\cite{yashin:2006a,kuksenok:2007a,yashin2007theoretical,kuksenok:2014b}. Here we will focus on simulating a simple but extensively-studied model, a thin elastic sheet driven by changes in its equilibrium metric. We will present an efficient semi-implicit time-stepping algorithm for the case of overdamped dynamics, a representative case with the same form of
numerical stiffness as more detailed fluid-elastic and fluid-poroelastic models. The approach is
also useful in other applications
where the dynamics of thin elastic sheets are important, such as
the self-assembly of thin sheets under magnetic forces \cite{boncheva2005magnetic,Alben2007saf}, and the
the rolling of actuated bilayers \cite{alben2011edge,Alben2015bending}.

A related problem is simulating the dynamics of fluid membrane vesicles with surface tension \cite{veerapaneni2011fast}. Here a similar time-step constraint arises for bending forces, while the thin elastic sheets considered in the present work also have stiffness due to stretching forces. Like \cite{veerapaneni2011fast}, we develop a semi-implicit time-stepping approach for computational efficiency, 
though our formulation differs due to the different mechanical forces.

In general, a semi-implicit (or implicit-explicit) time discretization writes some of the
terms (typically those with the highest spatial derivatives) implicitly, to improve time-step
constraints for stability \cite{rosenbrock1963some,gottlieb1977numerical,Hou1994removing,desbrun1999interactive,eberhardt2000implicit,Hou2001boundary,selle2008mass,Alben2008implicit,alben2009simulating}. The implicit terms are typically linear in the unknowns at the
current time step, so they can be solved directly at each time step, avoiding some of the complication and computational expense of nonlinear iterative solvers (e.g. Newton-type methods) in fully implicit discretizations \cite{chen2018physical}. If the linearized implicit term is sufficiently large in comparison
to the explicit terms, the semi-implicit method may be stable for a wide range of time
steps. For nonlinear PDEs, a somewhat empirical 
approach to formulating schemes, based on analogies with time-stepping for simpler linear PDEs, is often necessary.


In this work we begin with Seung and Nelson's discretization of elastic sheets by bending
and stretching springs on a triangular lattice \cite{Seung:1988}. We use the approach of \cite{desbrun1999interactive,selle2008mass} to
split the stretching force into an implicit linear term corresponding to zero-rest-length springs,
and a nonlinear remainder. The implicit stretching term is proportional to a discrete Laplacian matrix multiplying the
current sheet position. To stabilize the bending force, we add an implicit bending term that is proportional to a
discrete biharmonic matrix multiplying the current sheet position. The resulting method appears to be stable for all time steps. 
To allow for general elastic constants, we formulate a finite-difference discretization of the elastic energy
with the analogous semi-implicit approach. We validate and compare the methods on test problems with
internal in-plane stretching forces (nontrivial equilibrium metrics) and study
the effects of basic physical parameters on the sheets' dynamics.



\section{Elastic sheet \label{sec:Models}}

We consider a thin sheet or bilayer that undergoes large time-dependent deformations due to internal
forces (from a prescribed, time- and space-varying reference metric). We assume the sheet obeys linear (Hookean) elasticity, 
but that the midsurface (the set of points located midway through the sheet in the thickness direction) can be an arbitrary smooth 
time-dependent surface, so elastic forces depend nonlinearly on its position.
The extension of the Kirchhoff-Love (and F\"{o}ppl-von-K\'{a}rm\'{a}n) models of elastic plates to nonflat reference
metrics and/or large deformations with smooth midsurfaces has been called the Koiter shell theory 
\cite{koiter1966nonlinear,ciarlet2000modele,vetter2013subdivision} or the theory
of non-Euclidean plates \cite{efrati:2007a,efrati:2009a}. Here we mainly follow the latter's notation.

The elastic energy involves stretching and bending energy terms determined by the position of the body, 
$\mathbf{r}(\mathbf{x})$. Here $\mathbf{r}$ lies in $\mathbb{R}^3$, as does the material coordinate
$\mathbf{x} = (x_1, x_2, x_3)$. In the classical situation, the sheet has a zero-energy flat state
$\mathbf{r}(\mathbf{x}) = \mathbf{x}$, where $x_3$ lies in the interval $[-h/2, h/2]$ ($h$ is the
sheet thickness, much smaller than the other dimensions), 
and $x_1$ and $x_2$ lie in a planar region, the same for each $x_3$. 
For a small line of material $\mathbf{\Delta x} = \tilde{\mathbf{x}} - \mathbf{x}$ connecting two material 
points $\tilde{\mathbf{x}}$ and $\mathbf{x}$, its squared length on the undeformed surface
is $dl^2 = \Delta x_i \Delta x_i$ (with summation over repeated indices). Denote the squared length on the deformed surface $\mathbf{r}(\mathbf{x})$ by $dl'^2$. 
Using the Taylor series of $\mathbf{r}(\mathbf{x})$ up to first derivatives we have
\begin{equation}
dl'^2 - dl^2 = 2 \epsilon_{ij} \Delta x_i  \Delta x_j.
\end{equation}
\nn Here $\epsilon_{ij}$ is the strain tensor,
\begin{equation}
\epsilon_{ij} = \frac{1}{2} \left(g_{ij} - \delta_{ij}\right). \label{epsilon}
\end{equation}
\nn where 
\begin{equation}
g_{ij} = \frac{\partial r_k}{\partial x_i} \frac{\partial r_k}{\partial x_j} \label{g}
\end{equation}
\nn is the metric tensor. For curved shells or active materials, the rest state may be curved and/or time-varying, in which
case the reference metric, $\delta_{ij}$ in (\ref{epsilon}), becomes $\bar{g}_{ij}(\mathbf{x},t)$, a time- and space-varying
function determined for example by chemical activity \cite{yoshida:1996a,yashin:2006a,klein:2007a}:
\begin{equation}
\epsilon_{ij} = \frac{1}{2} \left(g_{ij} - \bar{g}_{ij}(\mathbf{x},t)\right). \label{epsilon1}
\end{equation}
 The reference metric is assumed to take the
form
\begin{equation}
\bar{g} = \begin{pmatrix}
\bar{g}_{11} & \bar{g}_{12} &\hfill 0   \cr
\bar{g}_{21} & \bar{g}_{22} &\hfill 0   \cr
0 & 0 & \hfill 1 
\end{pmatrix}
\end{equation}
\nn with upper 2-by-2 reference metric $\bar{g}_{\alpha \beta}$. The dependence
on out-of-plane components ($\bar{g}_{\alpha 3}, \bar{g}_{3\alpha}$) is trivial, so shearing though the plate thickness is not imposed and expansion/contraction in the thickness direction occurs only passively, due to the Poisson ratio effect. 
The components of the metric $g_{i 3}$ for $i = 1, 2, 3$ are determined by the Kirchhoff-Love assumptions
of no shearing in planes through the thickness, and no stress in the thickness direction. The result is
$g_{\alpha 3} = g_{3\alpha} = 0$ for $\alpha = 1,2$, and $g_{33} = 1$ \cite{efrati:2009a}. 

We may write the energy in terms of the midsurface deformation by expanding $g_{\alpha \beta}$ about the sheet midsurface $x_3 = 0$.
We obtain, at leading order,
its thickness-average $a_{\alpha\beta}$ and its thickness-gradient \cite{efrati:2009a}:
\begin{equation}
g_{\alpha\beta} = a_{\alpha\beta} - 2 x_3 b_{\alpha\beta} + O(h^2). \label{ab}
\end{equation}
\nn Here $a_{\alpha\beta}$ is the upper 2-by-2 part of the metric tensor (\ref{g}) evaluated at the sheet midsurface, $x_3 = 0$.  
The thickness-gradient is written in terms of $b_{\alpha\beta}$, the second fundamental form
\begin{equation}
b_{\alpha\beta} = \frac{\partial^2 r_k}{\partial x_\alpha \partial x_\beta} n_k
\end{equation}
\nn also evaluated at the sheet midsurface, $x_3 = 0$, with $n_k$ the components of the 
midsurface unit normal vector $\mathbf{n}$. 
For the reference metric $\bar{g}$, we write $\bar{a}$ and $\bar{b}$ for 
the corresponding terms in the expansion about the midsurface.

For an isotropic sheet with Young's modulus $E$, Poisson ratio $\nu$, and
thickness $h$, the elastic energy per unit volume is 
\begin{equation}
w = \frac{1}{2}\bar{A}^{\alpha\beta\gamma\delta}\epsilon_{\alpha\beta} \epsilon_{\gamma\delta},
\label{w}
\end{equation}
\nn a quadratic function of the in-plane
components of the strain tensor in (\ref{epsilon1})
with elasticity tensor
\begin{equation}
\bar{A}^{\alpha\beta\gamma\delta} = \frac{E}{1+\nu} \left(\frac{\nu}{1-\nu}\bar{g}^{\alpha\beta}\bar{g}^{\gamma\delta} +
\bar{g}^{\alpha\gamma}\bar{g}^{\beta\delta}\right) \label{barA}
\end{equation}
\nn where the entries of $\bar{g}^{\alpha\beta}$ are those of $\bar{g}_{\alpha\beta}^{-1}$. 
Integrating $w$ over the sheet thickness,
the energy per unit midsurface area is
\begin{equation}
w_{2D} = \int_{-h/2}^{h/2} w dx_3 = w_s + w_b + h.o.t.,
\label{w2D}
\end{equation}
\nn a sum of stretching energy per unit area
\begin{equation}
w_s = \frac{h}{8}A^{\alpha\beta\gamma\delta}\left(a_{\alpha\beta} - \bar{a}_{\alpha\beta}\right) 
\left( a_{\gamma\delta} - \bar{a}_{\gamma\delta}\right) \label{ws}
\end{equation}
\nn and bending energy per unit area
\begin{equation}
w_b = \frac{h^3}{24}A^{\alpha\beta\gamma\delta}\left(b_{\alpha\beta} - \bar{b}_{\alpha\beta}\right) 
\left( b_{\gamma\delta} - \bar{b}_{\gamma\delta}\right) . \label{wb}
\end{equation}
\nn in terms of the midsurface elasticity tensor
\begin{equation}
A^{\alpha\beta\gamma\delta} = \frac{E}{1+\nu} \left(\frac{\nu}{1-\nu}\bar{a}^{\alpha\beta}\bar{a}^{\gamma\delta} +
\bar{a}^{\alpha\gamma}\bar{a}^{\beta\delta}\right), \label{A}
\end{equation}
\nn and higher order terms in $h$.
The total elastic energy is, to leading order in $h$,
\begin{equation}
W = W_s + W_b \; , \quad W_s = \iint w_s \sqrt{|\bar{a}|} dx_1 dx_2 \;, \quad W_b =  \iint w_b \sqrt{|\bar{a}|} dx_1 dx_2. \label{W}
\end{equation}
\nn $W$ is a function of the midsurface configuration, $\mathbf{r}(x_1,x_2,0)$ and midsurface reference metric $\bar{a}$.

%




\section{Sheet dynamics \label{sec:dynamics}}

The sheet midplane evolves according a force balance equation,
where the elastic force per unit area $\mathbf{f}$ acting at a point on the midplane is
a sum of stretching and bending forces per unit area:
\begin{equation}
\mathbf{f} = \mathbf{f}_s + \mathbf{f}_b = \delta w_s/\delta \mathbf{r} + \delta w_b/\delta \mathbf{r}
\end{equation}
\nn given by taking the variation of $w_s + w_b$ with respect to $\mathbf{r}$. For a sheet moving in 
Stokes flow (i.e. at zero Reynolds number) the elastic forces would be balanced by external fluid forces which
depend linearly on the sheet velocity \cite{veerapaneni2011fast}:
\begin{equation}
\frac{\partial \mathbf{r}}{\partial t} = \mathcal{S}[\mathbf{f}_s + \mathbf{f}_b](\mathbf{r}). \label{Stokes}
\end{equation}
\nn where $\mathcal{S}$ is the Stokes operator
\begin{equation}
\mathcal{S}[\mathbf{f}](\mathbf{r}) = \int G(\mathbf{r},\mathbf{r'}) \mathbf{f}(\mathbf{r'}) dA(\mathbf{r'}),  \quad
G(\mathbf{r},\mathbf{r'}) = \frac{1}{8\pi\mu} \left(\frac{1}{\|\rho\|}\mathbf{I} + 
\frac{\rho\otimes\rho}{\|\rho\|^3} \right), \quad \rho \equiv \mathbf{r} - \mathbf{r'}.
\end{equation}
\nn In this work, we consider a simplification
of (\ref{Stokes}) which has similar numerical stiffness issues: overdamped dynamics, 
in which the Stokes operator is replaced with a multiple of the identity:
\begin{equation}
\mu \frac{\partial\mathbf{r}}{\partial t} = \mathbf{f}. \label{overdamped}
\end{equation}
\nn Here $\mu$ is a parameter that can be used to model the effects of internal and external damping. The equation can also be used to identify equilibria as a gradient descent method \cite{nocedal2006numerical} with time playing a nonphysical role. Extensions of
(\ref{overdamped}) including the inertia of the sheet or a surrounding fluid will add a
dependence on $\partial^2 \mathbf{r}/\partial t^2$. When discretized, the
$~\Delta t^{-2}$ dependence of such terms
will improve the time-step constraint compared to the overdamped problem, for
explicit schemes \cite{hairer1987solving}. In this work, we develop a semi-implicit
approach for (\ref{overdamped}), which should also apply with more detailed
forms of internal and external damping/forcing, e.g. from a fluid.
In the subsequent computational results we nondimensionalize the sheet lengths by the radius or half-width $R$ (e.g. for hexagonal and
square sheets), energy by the bending energy scale $Eh^3/12$, and time by the
period of $\bar{g}_{\alpha\beta}(\mathbf{x},t)$ (assumed to be time-periodic). These choices define
dimensionless versions of all the parameters (e.g. $\mu$).

%
%

\section{Triangular lattice with stretching and bending springs \label{sec:TriLattice}}

\begin{figure} [h]
           \begin{center}
           \begin{tabular}{c}
               \includegraphics[width=4in]{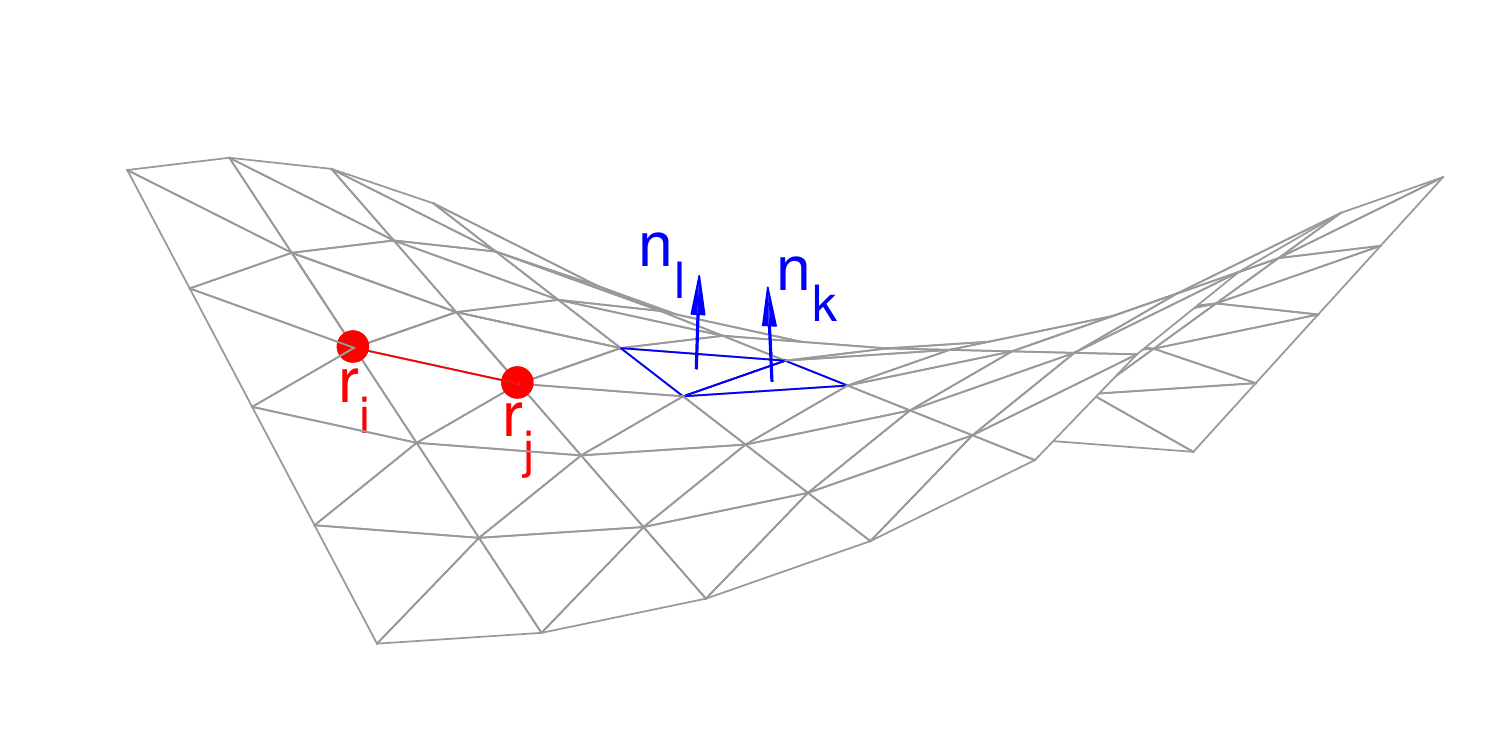} \\
           \vspace{-.25in} \hspace{-.25in}
           \end{tabular}
          \caption{\footnotesize Triangular lattice mesh with examples of
elements in the elastic energy: adjacent vertices $\mathbf{r}_i$ and $\mathbf{r}_j$ (red) and
adjacent face normals $\mathbf{n}_k$ and $\mathbf{n}_l$ (blue).
 \label{fig:TriangularLatticeSchematic}}
           \end{center}
         \vspace{-.10in}
        \end{figure}

We first consider a simple model of an elastic sheet with material points connected by an equilateral triangular lattice mesh. Nearest neighbor points are connected by Hookean springs and the total stretching energy is a sum of
the squares of nearest neighbor distances minus the local spring rest length $d_{ij}$:
\begin{equation}
U_s = \frac{K_s}{2} \sum_{i,j} (\|\mathbf{r}_i - \mathbf{r}_j\| - d_{ij})^2, \label{Us}
\end{equation}
\nn with $K_s$ a stretching stiffness constant. 
A bending energy is applied to adjacent triangular faces based on the angles between the normals to
the faces. The total bending energy is a sum over nearest neighbor pairs with bending stiffness 
constant $K_b$:
\begin{equation}
U_b = \frac{K_b}{2} \sum_{k,l}  \|\mathbf{n}_k - \mathbf{n}_l\|^2 = K_b \sum_{k,l} 1 - \mathbf{n}_k \cdot \mathbf{n}_l. \label{Ub}
\end{equation}
\nn Seung and Nelson used this model to study buckling due to defects in elastic membranes \cite{Seung:1988}, and it was used by 
many other groups to
study other deformations of thin sheets and shells due to defects and/or external forces 
\cite{gompper1997fluctuations,lobkovsky1997properties,lidmar2003virus,vliegenthart2006forced,katifori2010foldable,couturier2013folding,funkhouser2013topological,Alben2015bending,wan2017thermal},
as well as polymerized and fluid membranes \cite{nelson2004statistical}.

Seung and Nelson showed that for a lattice with $d_{ij} \equiv d$, a constant, as $d$ tends to 0 the stretching energy $U_s$ tends to that of an isotropic thin sheet with stretching rigidity $Eh = 2K_s/\sqrt{3}$ and Poisson ratio $\nu = 1/3$. The continuum limit of the bending energy contains two terms, one
proportional to the mean curvature squared and the other proportional to the Gaussian curvature. The term involving
mean curvature tends to that of an isotropic thin sheet with 
bending rigidity $Eh^3/12 (1-\nu^2) = \sqrt{3}K_b/2$. However, with this bending rigidity, 
the prefactor of the Gaussian curvature term is too large for $\nu > -1/3$ (too large by a factor of two at $\nu = 1/3$) \cite{schmidt2012universal}.
Nonetheless, for many problems, the Gaussian curvature term plays a negligible role because
it can be integrated to yield only boundary terms.  For closed shells, the Gaussian curvature term integrates to a constant and thus does not affect the elastic forces \cite{lidmar2003virus}. For open sheets (with boundaries), the equilibrium shape could be insensitive to the boundary shape or boundary conditions, particularly if the external or internal actuation is not localized at the boundary. In this work we will compare the model to a finite-difference discretization with $\nu = 1/3$
in two cases, and find a very small difference in a case of static actuation (also found in another situation by
\cite{didonna2002scaling}), and a somewhat larger difference in a case of dynamic actuation.

\begin{figure} [h]
           \begin{center}
           \begin{tabular}{c}
               \includegraphics[width=6.5in]{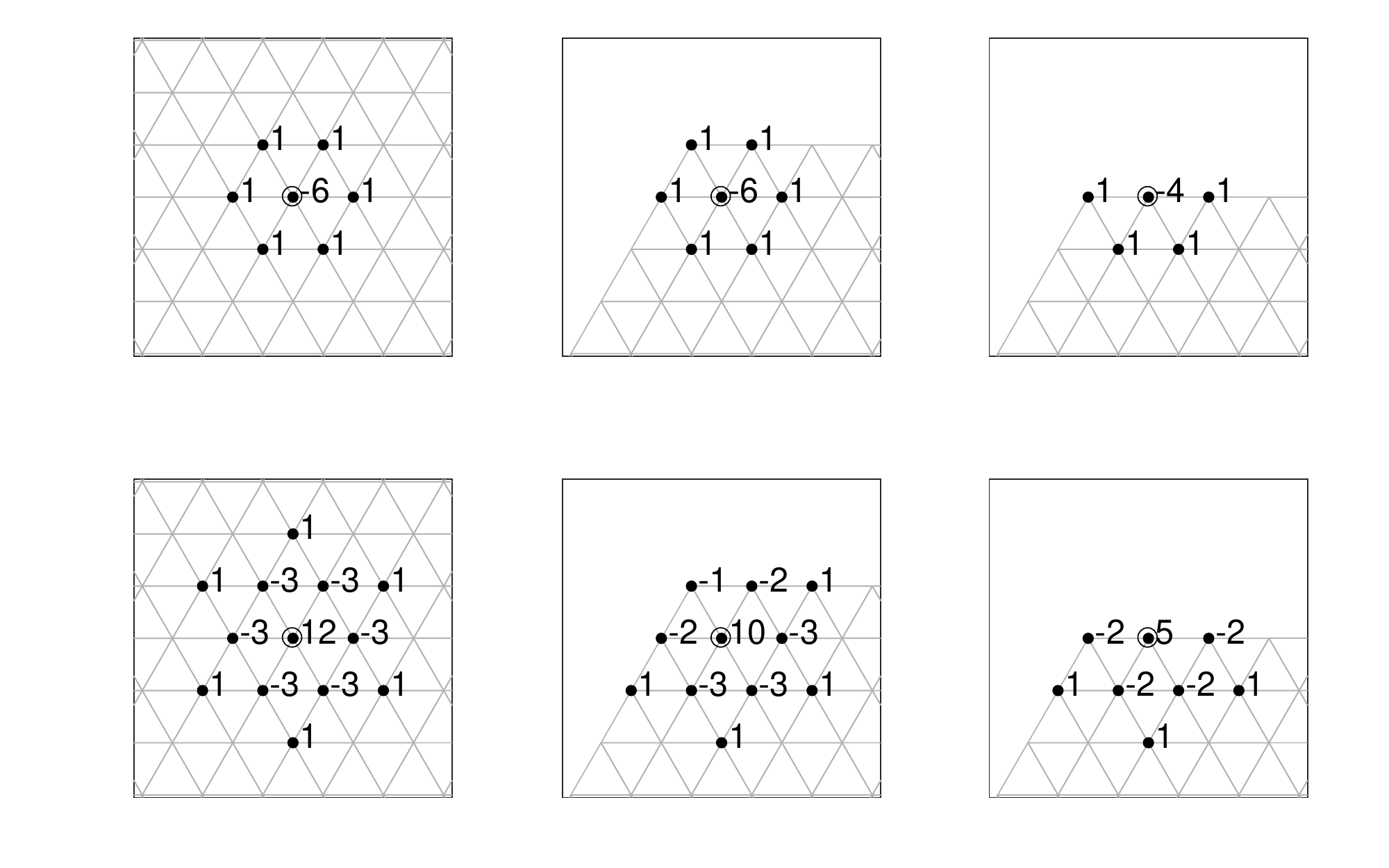} \\
           \vspace{-.25in} \hspace{-.25in}
           \end{tabular}
          \caption{\footnotesize Examples of the stencils corresponding to a discrete Laplacian operator with free edge boundary conditions (one of the diagonal blocks of $\mathbf{L}$ defined below (\ref{GradUs1})) (top row), and $\tilde{\Delta}^2_{x_1,x_2}$, a discrete biharmonic operator with free edge boundary conditions
(bottom row), at different mesh points (circled) away from and near a boundary on the triangular lattice.
 \label{fig:BiharmStencilFig}}
           \end{center}
         \vspace{-.10in}
        \end{figure}

The triangular lattice sheet is useful computationally because it has a simple expression for the energy and motivates our semi-implicit approach for the more general finite difference discretization given subsequently. The elastic force on the triangular lattice is obtained
by taking gradients of (\ref{Us}) and (\ref{Ub}) with respect to vertex coordinates $\left\{\mathbf{r}_i\right\}$.
The gradient of (\ref{Us}) with respect to $\mathbf{r}_i$ is
\begin{equation}
\nabla_{\mathbf{r}_i} U_s = K_s \sum_{\displaystyle j \in \mbox{nhbrs}(i)} (\|\mathbf{r}_i - \mathbf{r}_j\| - d_{ij}) \frac{(\mathbf{r}_i - \mathbf{r}_j)}{\|\mathbf{r}_i - \mathbf{r}_j\|}. \label{GradUs}
\end{equation}
\nn where nhbrs$(i)$ is the set of vertex neighbors to $i$. Following \cite{desbrun1999interactive,selle2008mass} 
we write the summand as a linear term plus a term with constant magnitude,
\begin{equation}
\nabla_{\mathbf{r}_i} U_s = K_s \sum_{\displaystyle j \in \mbox{nhbrs}(i)} \mathbf{r}_i - \mathbf{r}_j - d_{ij} \frac{(\mathbf{r}_i - \mathbf{r}_j)}{\|\mathbf{r}_i - \mathbf{r}_j\|} . \label{GradUs1}
\end{equation}
\nn To write the algorithms we define 
\begin{equation}
\bm{r} = \left[\bm{r}_x^\intercal, \bm{r}_y^\intercal, \bm{r}_z^\intercal\right]^\intercal
\end{equation} as the vector of 3$N$ vertex coordinates, with
$\bm{r}_x$, $\bm{r}_y$, and $\bm{r}_z$ the $N$-vectors of 
$x$-, $y$-, and $z$-coordinates. Thus $\mathbf{r}_i = \left[
\left(\bm{r}_x\right)_i, \left(\bm{r}_y\right)_i, \left(\bm{r}_z\right)_i \right]^\intercal$.
Also note that the italicized $\bm{r} \in \mathbb{R}^{3N}$ is different from the position function $\mathbf{r}(\mathbf{x})$ 
and a vertex on the discretized surface $\mathbf{r}_i$, both of which take values in $\mathbb{R}^{3}$.
It turns out that the linear term in (\ref{GradUs1}) can be written as the
product of $K_s$ and a block diagonal matrix $\mathbf{L}$ with $\bm{r}$, where $\mathbf{L}$ has
three blocks (along the diagonal), each of which is a discretized Laplacian on the triangular mesh with free-edge boundary 
conditions (see examples of stencils in figure \ref{fig:BiharmStencilFig}, top row), a negative semidefinite matrix.
Each block multiplies $\bm{r}_x$, $\bm{r}_y$, and $\bm{r}_z$, respectively.
We treat the linear term implicitly and the constant-magnitude term explicitly. Collecting the
terms (\ref{GradUs1}) for all vertices $i$, we obtain the total stretching force. A semi-implicit first-order temporal
discretization of (\ref{overdamped}) with stretching forces only is:
\begin{equation}
\mu A_p \frac{\bm{r}^{n+1} - \bm{r}^{n}}{\Delta t} = K_s \mathbf{L} \bm{r}^{n+1} +\mathbf{f}_{SE}(\bm{r}^n) \; , \quad \left[\mathbf{f}_{SE}(\bm{r})_i, \mathbf{f}_{SE}(\bm{r})_{N+i}, \mathbf{f}_{SE}(\bm{r})_{2N+i}\right]^\intercal
 \equiv - K_s \sum_{\displaystyle j \in \mbox{nhbrs}(i)}  d_{ij} \frac{(\mathbf{r}_i - \mathbf{r}_j)}{\|\mathbf{r}_i - \mathbf{r}_j\|}. \label{TimeStep}
\end{equation}
\nn Here $A_p = n_p\sqrt{3}d^2/12$ is the area per point on the undeformed lattice, with $n_p$ 
the number of triangles of which the point is a vertex, 6 for interior points and fewer for boundary points.
$\mathbf{f}_{SE}(\bm{r})$ is the nonlinear term in (\ref{GradUs1}), with 3$N$ entries, given
on the right side of (\ref{TimeStep}) for $i = 1, \ldots, N$.
 
Now assume the spring rest lengths are bounded for all time: $d_{ij} \leq \bar{d}$, a constant, and each
vertex has at most $p$ neighbors (6 for the triangular lattice). Rearranging
(\ref{TimeStep}) and using the boundedness of $\mathbf{f}_{SE}(\bm{r})$, we have an upper bound at time step $n+1$:
\begin{eqnarray}
\left\|\bm{r}^{n+1}\right\| &\leq& \left\|(\mathbf{I} - \Delta t K_s \mathbf{L}/(\mu A_p))^{-1} \bm{r}^n \right\| + \left\| (\mathbf{I} - \Delta t K_s \mathbf{L}/(\mu A_p))^{-1}\mathbf{f}_{SE}(\bm{r}^n) \right\| \\
&\leq& \left\| \bm{r}^n \right\| + K_s p\bar{d}  \\
&\leq& \left\| \bm{r}^0 \right\| + (n+1) K_s p\bar{d} .
\end{eqnarray}
So $\|\bm{r}^{n+1}\|$ grows at most linearly in time with this discretization. Empirically, the iteration appears to be bounded for all time steps for spring rest lengths $d_{ij}$ that are bounded in time. For comparison,
a forward Euler discretization of (\ref{overdamped}) with stretching forces only results in 
\begin{equation}
\bm{r}^{n+1} = \left(\mathbf{I} + \Delta t K_s \mathbf{L}/(\mu A_p)\right) \bm{r}^{n} + \frac{\Delta t }{\mu A_p} \mathbf{f}_{SE}(\bm{r}^n). \label{FEStretch}
\end{equation}
\nn Neglecting the rightmost term in (\ref{FEStretch}), we have a 2D diffusion equation, and stability is possible only when the largest eigenvalue of $\mathbf{I} + \Delta t K_s \mathbf{L}/(\mu A_p)$ is bounded in magnitude by 1. Since the eigenvalues of $\mathbf{L} \sim 1/\Delta x^2$ for lattice spacing $\Delta x$, this requires $\Delta t < C_s \mu A_p \Delta x^2 / K_s$ for a constant $C_s$.

The gradient of the bending energy (\ref{Ub}) with respect to a lattice vertex $\mathbf{r}_i$ is
\begin{equation}
\nabla_{\mathbf{r}_i} U_b = K_b \sum_{k,l} \sin{\theta_{kl}}\nabla_{\mathbf{r}_i}\theta_{kl}, \label{GradUb}
\end{equation}
\nn using $\mathbf{n}_k \cdot \mathbf{n}_l = \cos{\theta_{kl}}$. The dihedral angle $\theta_{kl}$ depends on the four points in the union of the neighboring triangles $k$ and $l$ (see figure \ref{fig:TriangularLatticeSchematic}). Two of
these points are the endpoints of the edge shared by the triangles.
At each of the other two points, $\nabla_{\mathbf{r}_i}\theta_{kl}$ is directed along 
the normal to the triangle in which it lies, with magnitude equal to the reciprocal of its distance from the
shared edge. At the endpoints of the shared edge, $\nabla_{\mathbf{r}_i}\theta_{kl}$ can be
found by requiring that the net force and torque due to $\theta_{kl}$ is zero, which gives
six equations (for the three components of net force and torque) in six unknowns (the forces
on the two endpoints of the shared edge). The bending force $\mathbf{f}_{B}$ is a 3$N$-vector with components
\begin{equation}
\left[\mathbf{f}_{B}(\bm{r})_i, \mathbf{f}_{B}(\bm{r})_{N+i}, \mathbf{f}_{B}(\bm{r})_{2N+i}\right]^\intercal
 \equiv -\nabla_{\mathbf{r}_i} U_b, \; i = 1, \ldots, N,
\end{equation}
\nn similar to $\mathbf{f}_{SE}$ in (\ref{TimeStep}). 
Our linearized approximation to the bending force is similar to that of \cite{veerapaneni2011fast}. They write the terms with the highest spatial derivatives in the form $\mathcal{B}(\bm{r}^n) \bm{r}^{n+1}$. Here $\mathcal{B}$ involves fourth derivatives with prefactors that include lower-order derivatives extrapolated to time step $n+1$ from previous time steps. In fact, we use a simpler expression: $\mathbf{B} \bm{r}^{n+1}$, where $\mathbf{B}$ is a block diagonal matrix with each of the three blocks
equal to 
$D \tilde{\Delta}^2_{x_1,x_2}$. Here $D$ is the bending modulus ($Eh^3/(12(1-\nu^2))$ in dimensional form, $1/(1-\nu^2)$ in dimensionless form) and $\tilde{\Delta}^2_{x_1,x_2}$ is the discretized biharmonic operator on the triangular lattice in the orthogonal material coordinates $x_1$ and $x_2$.
If in-plane strain (shearing and dilation) are not very large,
$x_1$ and $x_2$ are close to orthogonal arclength coordinates $s_1$, $s_2$ along the midplane surface. For
any surface $\mathbf{X}(s_1, s_2)$ parametrized by orthogonal arclength coordinates $s_1$, $s_2$ we can write
\begin{equation}
\Delta^2_{s_1,s_2} \mathbf{X}(s_1, s_2) = \Delta_{s_1,s_2} (\kappa_1+\kappa_2)\hat{\mathbf{n}} + N. \label{Biharm12}
\end{equation}
\nn where $\kappa_1+\kappa_2$ is twice the mean curvature and 
$N =(\kappa_1+\kappa_2)\Delta_{s_1,s_2}\hat{\mathbf{n}}$ involves derivatives
of $\mathbf{X}$ that are of lower order than those in the first term on the right hand side. 
The highest-derivative term in the continuum bending force is also the first 
term on the right hand side of (\ref{Biharm12}) when the equilibrium metric is
the identity (see \cite{weiner1978problem,veerapaneni2011fast}). Thus the
left hand side of (\ref{Biharm12}) is a reasonable linear (and constant-coefficient) 
approximation to the bending force. A more accurate linear approximation
to the bending operator could include corrections that take into account the nontrivial
inverse equilibrium metric ($\bar{g}^{\alpha\beta}$ in (\ref{A})) and in-plane strain, e.g. by including 
nonuniform prefactors extrapolated from previous time steps. 
We find however that the constant-coefficient biharmonic is a sufficiently good approximation in the
sense that it damps out spurious mesh-scale bending oscillations (as occurs with a fully explicit bending
term) up to large-amplitude variations in the reference metric, 
$\approx 0.3$ in terms of an amplitude parameter $A$ defined in (\ref{eta1})--(\ref{eta3}), below. With
larger variations in the reference metric, the position vector remains bounded in time, but there is very large deformation and self-intersection even for the case of purely planar deformations, without bending forces. 
We explain how the discrete biharmonic operator $\tilde{\Delta}^2_{x_1,x_2}$ is calculated in appendix \ref{sec:BendingForce}.

Semi-implicit (or implicit-explicit) schemes for cloth animation have sometimes left bending
forces explicit, when they are much smaller than stretching forces \cite{eberhardt2000implicit}.
Considering a forward Euler discretization of (\ref{overdamped}) with bending forces only, and
approximating the bending force by $\mathbf{B} \bm{r}^{n}$, stability requires
$\Delta t < C_b \mu  A_p \Delta x^4 / K_b$ for a constant $C_b$, since the eigenvalues of $\mathbf{B} \sim 1/\Delta x^4$ for lattice spacing $\Delta x$. The ratio of bending to stretching time step constraints $\sim K_s \Delta x^2 /K_b \sim \Delta x^2 /h^2$, the square of the ratio of lattice spacing to sheet thickness. For the
parameters used in this work, the stretching time step constraint is typically smaller, but not much
smaller than the bending time constraint, so a semi-implicit approach needs to address both terms.

Our first-order semi-implicit discretization for sheet dynamics with both stretching and bending forces is:
\begin{equation}
\mu A_p \frac{\bm{r}^{n+1} - \bm{r}^{n}}{\Delta t} = K_s \mathbf{L} \bm{r}^{n+1} +\mathbf{f}_{SE}(\bm{r}^n) + \mathbf{B} \bm{r}^{n+1} -\mathbf{B} \bm{r}^{n} +
\mathbf{f}_{B}(\bm{r}^n) . \label{SB1}
\end{equation}
\nn The last two terms on the right hand side approximately cancel in the highest-derivative term, leaving the 
implicit bending term $\mathbf{B} \bm{r}^{n+1}$ as the dominant one.
The second-order version with uniform time stepping (an approximate backward differentiation formula) is:
\begin{eqnarray}
\mu A_p \frac{3\bm{r}^{n+1} - 4\bm{r}^{n} + \bm{r}^{n-1}}{2\Delta t} =  K_s \mathbf{L} \bm{r}^{n+1} + 2 \mathbf{f}_{SE}(\bm{r}^n) - \mathbf{f}_{SE}(\bm{r}^{n-1}) + \mathbf{B} \bm{r}^{n+1} -2\mathbf{B} \bm{r}^{n}+\mathbf{B} \bm{r}^{n-1} +2 \mathbf{f}_{B}(\bm{r}^n) - \mathbf{f}_{B}(\bm{r}^{n-1}). \label{SB2}
\end{eqnarray}
\nn For test problems that model the deformation and dynamics of active gel sheets driven by internal swelling 
\cite{yoshida:1996a,yashin:2006a,yashin2007theoretical,klein:2007a}, we construct a uniform triangular
lattice with spacing $d$. We then set the dilatation factor
$d_{ij}/d \equiv \eta$ to correspond to one of three examples:
a static radial distribution, a unidirectional traveling wave,
or a radial traveling wave of isotropic dilation/contraction:
\begin{eqnarray}
\eta_1(x_1, x_2, t) &=& 1+A\cos\left({2\pi\left(k \sqrt{x_1^2 + x_2^2}\right)}\right) \label{eta1} \\
\eta_2(x_1, x_2, t) &=& 1+A\sin\left({2\pi\left(k x_1 - t\right)}\right) \label{eta2} \\
\eta_3(x_1, x_2, t) &=& 1+A\sin\left({2\pi\left(k \sqrt{x_1^2 + x_2^2} - t\right)}\right).  \label{eta3}
\end{eqnarray}
\nn The corresponding equilibrium metrics are
\begin{equation}
\bar{a}_{\alpha \beta}(x_1, x_2, t) = \eta^2(x_1, x_2, t) \delta_{\alpha \beta}, \label{bara}
\end{equation}
\nn and we take zero reference curvature ($\bar{b}_{\alpha \beta} = 0$) for simplicity. Before presenting results, we describe in the next section a more direct finite difference simulation that
allows for more general elastic parameters than the triangular lattice
model. However, the triangular lattice simulations 
have the advantage of remaining bounded in time for all time steps across wide ranges (several
orders of magnitude) of values for the parameters ($h$, $A$, $k$, $\mu$) with large mesh sizes (hexagonal
domains with $N$ = 3367 and 9241, for example). Simulations appear smooth
up to strain amplitudes $A \sim 0.3$. Above this value, sheet shapes become jagged and self-intersect, but remain
bounded in time. Hence the time step and lattice spacing are 
constrained only by the need to resolve the dynamics at a given parameter set. The method can also
be used to converge to static equilibria, in which case the time step becomes a step length for a gradient descent algorithm.
Here a large step length may be used initially to rapidly approach the neighborhood of an equilibrium, and then
a smaller step length allows for convergence. The resulting convergence is geometric 
(not superlinear) but generally quite fast,
and due to the simplicity of the formulation it is a good alternative to Newton and quasi-Newton methods in the static
case (as well as the dynamic case).


\section{Finite difference algorithm}

We now propose a second algorithm, inspired by that for the triangular lattice, 
but based on a finite difference discretization of the continuum elastic energy (\ref{W}), 
and which therefore allows the full range of values of $E$, $h$, and $\nu$ (that are physically
reasonable). We use a square grid with grid spacing $\Delta x$ for simplicity and define second-order accurate 
finite-difference operators:
\begin{equation}
D_{\alpha} \approx \partial_{\alpha} \; , \;  D_{\alpha\beta } \approx \partial^2_{\alpha \beta}\; , \; \alpha, \beta = \{1,2\}
\end{equation}
\nn In the energy (\ref{W}) we use
\begin{equation}
a_{\alpha\beta} \approx D_{\alpha}\bm{r}_x \odot D_{\beta}\bm{r}_x +
D_{\alpha}\bm{r}_y \odot D_{\beta}\bm{r}_y +
D_{\alpha}\bm{r}_z \odot D_{\beta}\bm{r}_z, \label{a}
\end{equation}
\nn where $\odot$ denotes a componentwise (Hadamard) product of two vectors. We use the analogous expression for $b_{\alpha\beta}$ and a trapezoidal-rule quadrature for the integrals in (\ref{W}). For equilibrium metrics in the
form (\ref{bara}) the discrete form of the stretching energy ($W_s$ in (\ref{W})) is
\begin{equation}
\tilde{W}_s = \frac{h}{8} A^{\alpha\beta\gamma\delta} 
 \left(\mathbf{q}\odot\eta^2\right)^\intercal \left[ \left( a_{\alpha\beta} - \bar{a}_{\alpha\beta}\right)\odot \left( a_{\gamma\delta} - \bar{a}_{\gamma\delta}\right) \right] \label{tildeWs}
\end{equation}
\nn where $\mathbf{q}$ is the vector of quadrature weights for the trapezoidal rule on the rectangular mesh,
and we use the usual summation rule for repeated indices. 
 We compute $\nabla_{\bm{r}_x} \tilde{W}_s$ by using the chain rule with (\ref{a}) and (\ref{tildeWs}):
\begin{equation}
\nabla_{\bm{r}_x} \tilde{W}_s = \frac{h}{4}D_\alpha^\intercal \left[A^{\alpha\beta\gamma\delta} 
 \left(\mathbf{q}\odot\eta^2 \odot \left( a_{\gamma\delta} - \bar{a}_{\gamma\delta}\right)\right) D_\beta \bm{r}_x) \right] + \frac{h}{4}D_\beta^\intercal \left[A^{\alpha\beta\gamma\delta} 
 \left(\mathbf{q}\odot\eta^2 \odot \left( a_{\gamma\delta} - \bar{a}_{\gamma\delta}\right)\right) D_\alpha \bm{r}_x) \right] \label{GradWs}
\end{equation}
\nn and the same expressions for $\nabla_{\bm{r}_y} \tilde{W}_s$ and $\nabla_{\bm{r}_z} \tilde{W}_s$, with $\bm{r}_x$ in (\ref{GradWs}) replaced by $\bm{r}_y$ and $\bm{r}_z$ respectively. Our linearized approximation to the stretching force, to compute $\bm{r}^{n+1}$ semi-implicitly, is
$\mathbf{M}_s^n \bm{r}^{n+1}$, where $\mathbf{M}_s^n$ is a block diagonal matrix with
three blocks, each given by
\begin{equation}
\frac{h}{4}D_\alpha^\intercal \left[A^{\alpha\beta\gamma\delta} 
 \left(\mathbf{q}\odot\eta^2 \odot  a_{\gamma\delta}^n  \right) D_\beta) \right] +
 \frac{h}{4}D_\beta^\intercal \left[A^{\alpha\beta\gamma\delta} 
 \left(\mathbf{q}\odot\eta^2 \odot  a_{\gamma\delta}^n  \right) D_\alpha) \right].
\end{equation}
\nn $\mathbf{M}_s^n \bm{r}^{n+1}$
is the discrete stretching force with zero reference metric, analogous to the discrete Laplacian on the triangular lattice, which
gives the stretching force with zero-rest-length springs. The use of $a_{\gamma\delta}^n$ instead of
$a_{\gamma\delta}^{n+1}$ makes 
$\mathbf{M}_s^n$ independent of $\bm{r}^{n+1}$. An extrapolation that is higher-order in time can also be used.

We also compute $\nabla_{\bm{r}} W_b$ using the chain rule, resulting in a similar (though somewhat lengthier) expression for the bending force. Our linearized approximation to the bending force is $\mathbf{M}_b \bm{r}^{n+1}$, where
$\mathbf{M}_b$ is the product of the bending modulus and the discrete biharmonic operator,
the same as for the triangular lattice but now on a rectangular mesh.

Unlike the triangular lattice algorithm, the finite difference algorithm is only stable for first-order time-stepping, i.e.
\begin{equation}
\mu A_p \frac{\bm{r}^{n+1} - \bm{r}^{n}}{\Delta t} = \mathbf{M}_s^n \bm{r}^{n+1} - \mathbf{M}_s^n \bm{r}^{n} 
-\left\{\nabla_{\bm{r}} \tilde{W}_s\right\}^{n}+ \mathbf{M}_b \bm{r}^{n+1} - \mathbf{M}_b \bm{r}^{n} 
-\left\{\nabla_{\bm{r}} \tilde{W}_b\right\}^{n}, \label{FD1}
\end{equation}
\nn the analogue
of (\ref{SB1}), with $A_p = \Delta x^2$ the area per point (multiplied by 1/2 at points along the sides and 1/4 at the corners). It is unstable for the second-order version,
\begin{eqnarray}
\mu A_p \frac{3\bm{r}^{n+1} - 4\bm{r}^{n} + \bm{r}^{n-1}}{2\Delta t} = &&  \mathbf{M}_s^{[n+1]} \bm{r}^{n+1} - 2 \mathbf{M}_s^{[n]} \bm{r}^{n} + \mathbf{M}_s^{[n-1]} \bm{r}^{n-1}
-2\left\{\nabla_{\bm{r}} \tilde{W}_s\right\}^{n}+ \left\{\nabla_{\bm{r}} \tilde{W}_s\right\}^{n-1}  \label{FD2a}\\
&+&   \mathbf{M}_b \bm{r}^{n+1} - 2 \mathbf{M}_b \bm{r}^{n} + \mathbf{M}_b \bm{r}^{n-1}
-2\left\{\nabla_{\bm{r}} \tilde{W}_b\right\}^{n}+ \left\{\nabla_{\bm{r}} \tilde{W}_b\right\}^{n-1}. \label{FD2b}
\end{eqnarray}
due to the extrapolated gradient terms (i.e.  (\ref{FD2a})--(\ref{FD2b}) becomes stable when the extrapolation reverts to first order for the
gradient terms in (\ref{FD2b})). The superscripts in brackets denote a second-order extrapolation to the indicated time step using
the two preceding time steps. We note that second-order accuracy can be obtained from the first order method
via Richardson extrapolation.



While the triangular lattice algorithm is essentially unconditionally stable, 
the first-order finite-difference algorithm is only stable up to moderately large time steps. However,
the semi-implicit operators do yield orders-of-magnitude improvements in the largest stable time steps 
compared to an explicit scheme (forward Euler). As an
example, we set $\eta = \eta_3$ in (\ref{eta3}), with $A = 0.1$, $h = 0.03$, $\mu = 1000$, and $\nu = 1/3$. In table \ref{TimeStepTable}
we give approximate values for the maximum stable time step, defined to be a time step such that
the sheet deflection remains bounded (below $10^8$ in maximum norm) up to $t = 5$. 
The sheet is a square of side length 2, with $n$ grid points in each direction.
As $n$ is increased from 33 to 66, the maximum stable time step decreases by a factor of 2 for the semi-implicit method, versus
a factor of 6 for the explicit method. Because the cost of solving the linear systems in
(\ref{SB1}), (\ref{SB2}), and (\ref{FD1}) is only slightly larger than the cost of 
the rest of the algorithm 
for the largest $n$ in table \ref{TimeStepTable}, the orders-of-magnitude difference in
time step translates to an orders-of-magnitude difference in overall computational cost for
these mesh sizes.

\begin{table}
\begin{center}
\vspace{1cm}
\begin{tabular}[t]{c|l|l|}
$n$ & $\Delta t^{SI}_{max}$ & $\Delta t^{FE}_{max}$ \\ \hline
33 &  0.04 & 1.8 $\times 10^{-4}$ \\ \hline
44 &  0.03 & 1.0 $\times 10^{-4}$ \\ \hline
55 & 0.025 & 5 $\times 10^{-5}$ \\ \hline
66 & 0.02 & 3 $\times 10^{-5}$ \\ \hline
\end{tabular}
\end{center}
\caption{\label{TimeStepTable} Approximate upper bounds on stable time step for the semi-implicit finite difference scheme (SI) compared
to forward Euler (FE), with different grid sizes $n$. The number of grid points is $n^2$ and the number of unknowns
is $3n^2$.}
\end{table}


\section{Results}


\begin{figure} [h]
           \begin{center}
           \begin{tabular}{c}
               \includegraphics[width=6.5in]{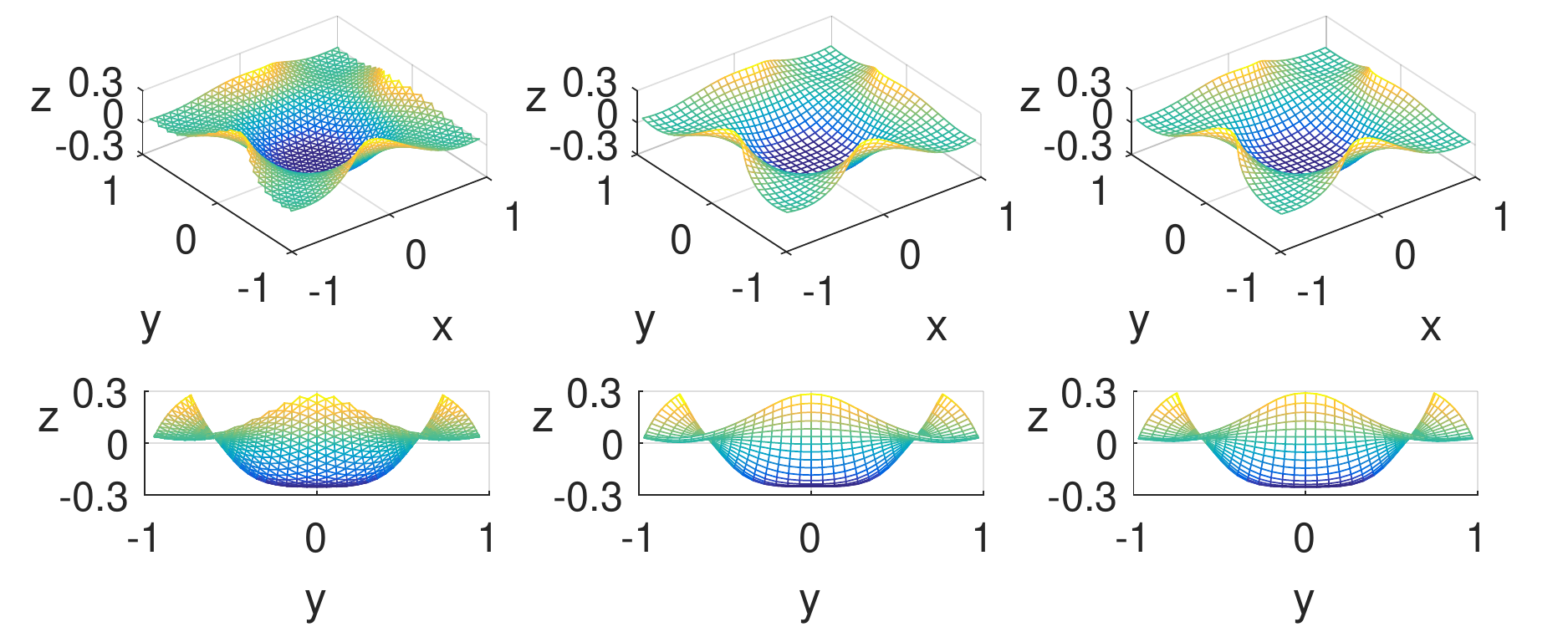} \\
           \vspace{-.25in} \hspace{-.25in}
           \end{tabular}
          \caption{\footnotesize Static equilibria for the triangular lattice algorithm (left), and 
the finite difference algorithms with
different Poisson ratios in one of the bending energy terms (center and right). The reference metrics are given
by $\eta =\eta_1$ in (\ref{eta1}), with physical parameters 
$A = 0.1$, $k = 1$, $h = 0.03$, $\mu = 1000$, and $\nu = 1/3$ or $-1/3$ in certain bending energy terms (see text).
 \label{fig:CompareTriRectStatic}}
           \end{center}
         \vspace{-.10in}
        \end{figure}

We now present a sequence of simulation results to display basic aspects of the algorithms and parameters. First,
we compare the triangular lattice and finite difference methods in two situations. The first is the
equilibrium sheet deformation with a nontrivial, but static (time-independent) reference metric given
by $\eta =\eta_1$ in (\ref{eta1}), with $A = 0.1$, $k = 1$, $h = 0.03$, $\mu = 1000$, and $\nu = 1/3$.
The sheets are initially nearly flat squares ($z = 0.02 (x_1^4 + x_2^4)$, $-1\leq x_1 = x, x_2 = y \leq 1$),  
and rapidly buckle into the shapes 
shown in figure \ref{fig:CompareTriRectStatic}. Three sheets are shown in oblique view (top row)
and side view (bottom row). The leftmost sheet is computed with the triangular lattice algorithm, with
jagged edges along one pair of sides. The center sheet is the result of the finite difference algorithm,
with a different Poisson ratio value in one of the bending energy terms, to match those of the triangular 
lattice algorithm. For equilibrium metrics (\ref{bara}) we may write (\ref{wb}) as
\begin{equation}
w_b = \frac{Eh^3}{24(1-\nu^2)}\eta^4\left( (b_{11} + b_{22})^2 -2(1-\nu) (b_{11} b_{22} - b_{12}^2) \right). \label{wb1}
\end{equation}
\nn In the stretching energy term (\ref{ws}) and the prefactor of the right hand side of (\ref{wb1}), we set
$\nu = 1/3$, but we set $\nu$ to -1/3 in the second appearance of $\nu$ (within $1-\nu$) on the right hand
side of (\ref{wb1}). This gives an elastic
energy consistent with that of the triangular lattice model \cite{schmidt2012universal}.
The rightmost sheet is also given by the finite difference algorithm but with $\nu = 1/3$ in all stretching
and bending energy terms. The deformations are very close in all three cases. The differences between
maximum and minimum $z$ values, $\Delta z$, are within 1\% for the center and leftmost sheets,
and the difference is about 2\% for the center and rightmost sheets. In this case at least, the
error in Gaussian bending rigidity in the triangular lattice algorithm has a modest effect,
as was found by \cite{didonna2002scaling} in a different problem.

\begin{figure} [h]
           \begin{center}
           \begin{tabular}{c}
               \includegraphics[width=6.5in]{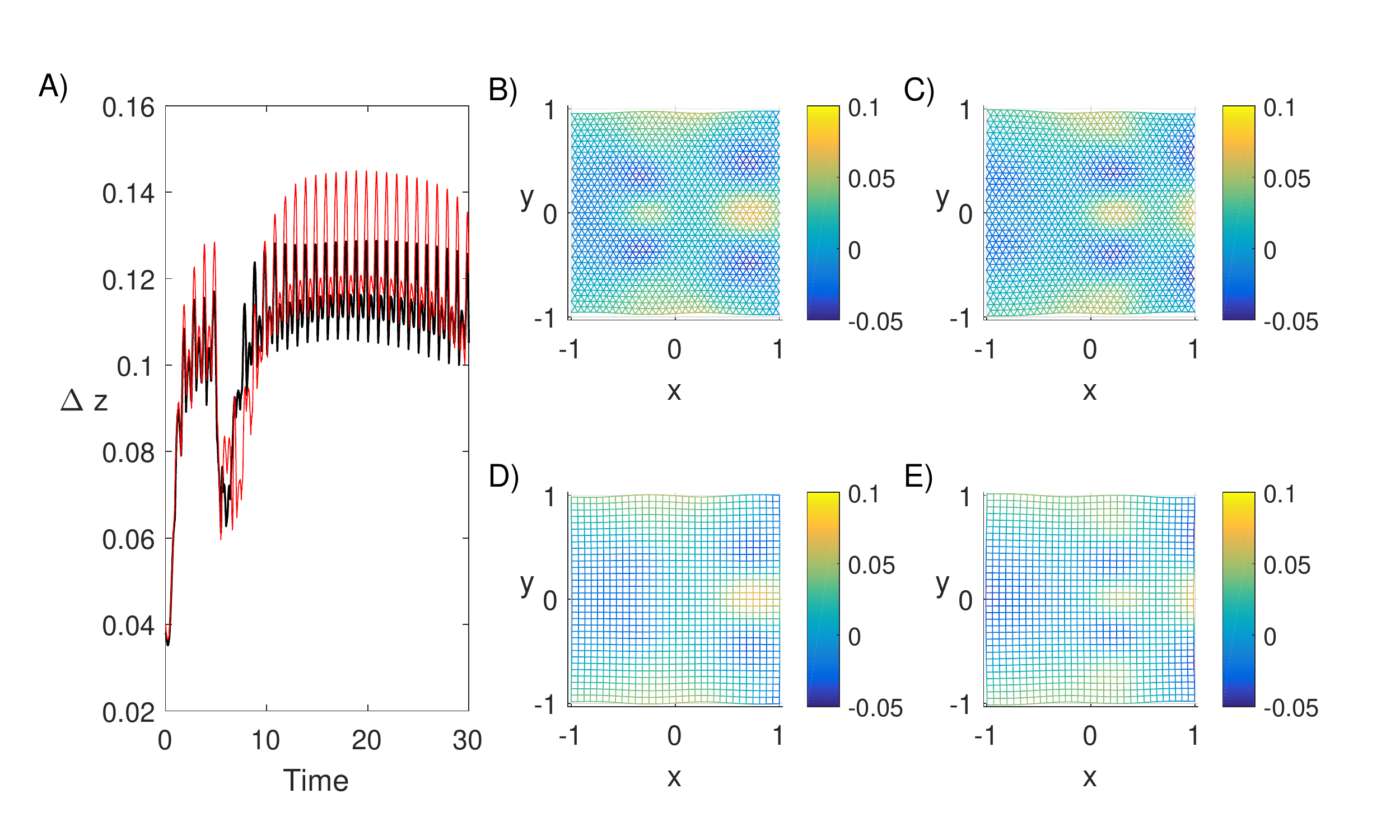} \\
           \vspace{-.25in} \hspace{-.25in} 
           \end{tabular}
          \caption{\footnotesize A comparison of sheet dynamics with the triangular lattice and
finite difference algorithms with reference metric factor $\eta =\eta_2$ in (\ref{eta2}), a unidirectional traveling wave, 
starting from a nearly flat sheet. The physical parameters are $A = 0.03$, $k = 1$, $h = 0.03$, $\mu = 1000$, and $\nu = 1/3$. A) A comparison of the $z$-deflection over time for the triangular lattice (black) and finite difference (red) algorithms. Distributions of $z$ deflection at time $t$ = 29.5 and 30 are shown in panels B and
C for the triangular lattice and D and E for the finite difference method, respectively.
 \label{fig:CompareTriRectWave}}
           \end{center}
         \vspace{-.10in}
        \end{figure}

We now consider an unsteady reference metric, $\eta =\eta_2$ in (\ref{eta2}), a unidirectional traveling wave, 
in both algorithms. We again take square sheets, with the same spatial grids and initial conditions as before, with (smaller) $A = 0.03$, $k = 1$, $h = 0.03$, $\mu = 1000$, and $\nu = 1/3$ (in all energy terms in both algorithms now, for simplicity). 
The time step $\Delta t = 0.005$. Figure \ref{fig:CompareTriRectWave} compares the dynamics. Panel A shows
$\Delta z$ versus time for the triangular lattice (black) and finite difference (red) algorithms. The dynamics
after buckling of the initial state are quite complex, and include the appearance of a higher buckling
mode during $6 \leq t \leq 8$. For both algorithms the solutions have a strong oscillatory component with the same frequency 
as the reference metric, as one would expect. 
The solutions have other components that evolve on much longer time scales (tens to hundreds
of periods), and even with modest deflections (here, about 7\% of the square side length),
the solutions can evolve in complicated ways over long time scales. The two algorithms show rough qualitative
agreement, though the discrepancy in deflection reaches 10--15\% at certain times in panel A. 
The $z$ deflection at $t$ = 29.5 and 30 is shown in panels B and C for
the triangular lattice and D and E for the finite difference method, respectively. Both algorithms show a similar distribution
of deflection on the right side of the sheets. On the left side, the triangular lattice has an additional peak (near $x = -0.25, y = 0$) that is not present in the finite difference method.

We have illustrated the behavior of the algorithms in two simple cases. We've seen close agreement in the final equilibrium
after buckling with a steady reference metric, and somewhat less agreement in a dynamical problem with a unidirectional traveling
wave metric. The disagreements may be attributed to the sensitivity of the dynamical problem to different prefactors in the Gaussian curvature term of the bending energy and to the different spatial discretizations. We now proceed to illustrate some other basic features of the buckling behavior and dynamics, with the triangular lattice approach only for brevity.

\begin{figure} [h]
           \begin{center}
           \begin{tabular}{c}
               \includegraphics[width=6in]{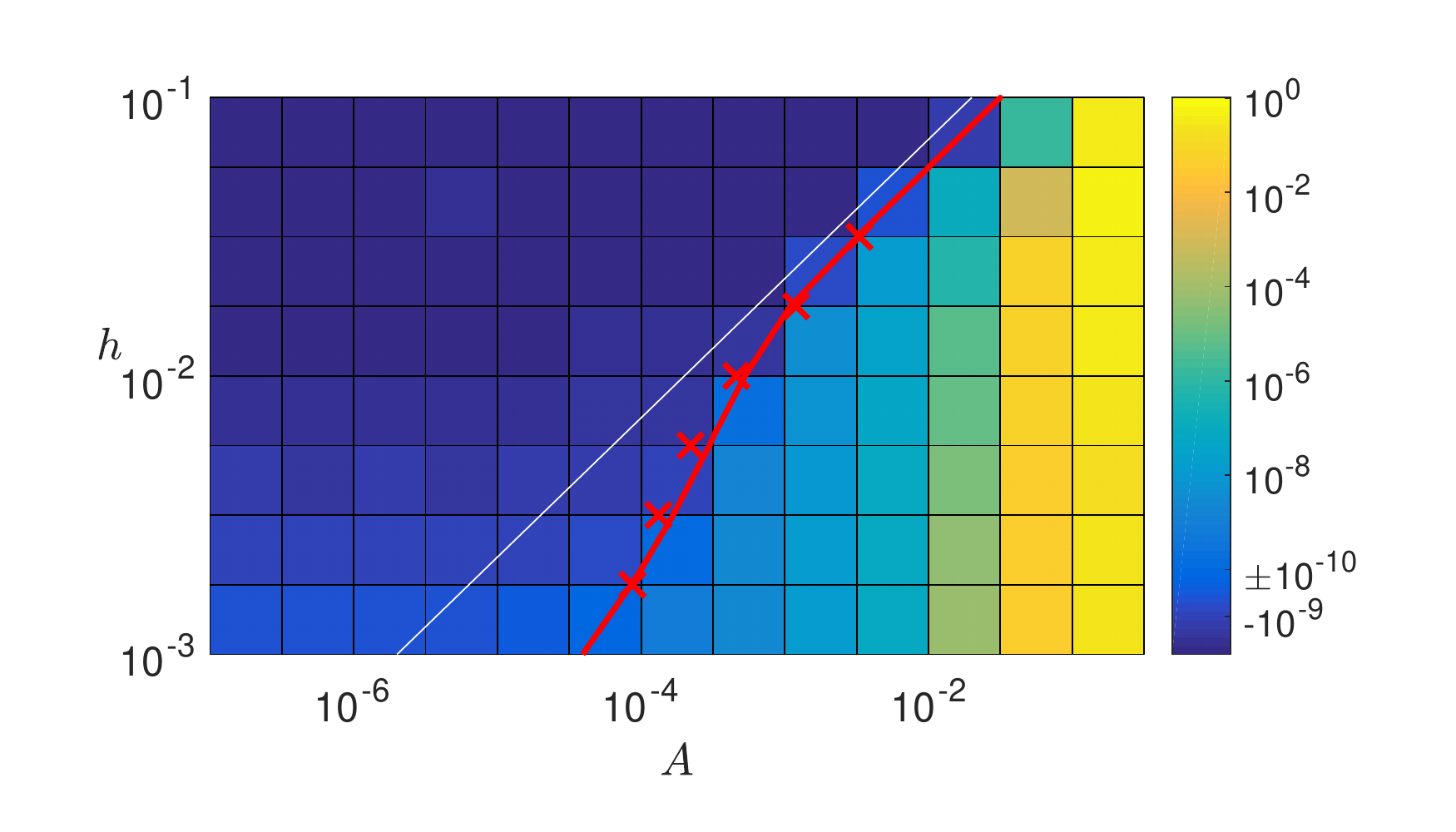} \\
           \vspace{-.25in} \hspace{-.25in}
           \end{tabular}
          \caption{\footnotesize Change in out-of-plane deflection from $t = 0$ to $t =2$ for a unit
hexagon with a small initial deflection $z = 10^{-8} x_1 x_2$, and imposed reference metric 
(\ref{eta1}) with $k$ = 1, varying metric factor amplitude $A$ (horizontal axis) and varying sheet thickness $h$ (vertical axis). 
The damping constant $\mu = 12.6$ and the mesh spacing $d$ = 1/33. The solid
red line shows the locus of zero change in deflection, essentially the ``buckling threshold."
The red crosses show, at several values of $h$, 
the values of $A$ on the buckling threshold with a finer mesh spacing, 1/66. The white line
shows the scaling $A \sim h^2$.
 \label{fig:BucklingGrowthFig}}
           \end{center}
         \vspace{-.10in}
        \end{figure}

If a static reference metric cannot be realized by a surface in $\mathbb{R}^3$, the sheet has nonzero stretching energy,
and in general becomes unstable to out-of-plane buckling for sufficiently small $h$ \cite{efrati:2009b}.
Intuitively, buckling allows a reduction in stretching energy at the expense of bending
energy. The relative cost of bending decreases with decreasing $h$, making buckling more favorable.
For the static metric with $\eta =\eta_1$ in (\ref{eta1}), we plot a computational estimate of the ``buckling threshold''---for 
various $A \in [10^{-5}, 10^{-1}]$, the values of $h$ at which buckling occurs. Here buckling
is defined by whether a small initial deflection ($z = 10^{-8} x_1 x_2$) grows after two time units, with a certain damping constant ($\mu = 12.6$). Changes in the time interval and damping constant have only a slight
effect on the buckling threshold, plotted as a red solid line for $d = 1/33$. The mesh spacing becomes
more important at smaller $h$, where buckling deformations may occur with a smaller wavelength due to
the decreased bending energy. To check the effect of mesh
refinement, we repeat the computations with $d = 1/66$ at selected points and obtain the red crosses. 
The white line shows the scaling $A \sim h^2$, 
which approximately matches the buckling threshold data presented
in \cite{efrati:2009b} for a different reference metric. Our data appear to follow this trend for $h \geq 0.01$. Deviations
at smaller $h$ are likely due to the finite mesh spacing.

\begin{figure} [h]
           \begin{center}
           \begin{tabular}{c}
               \includegraphics[width=6in]{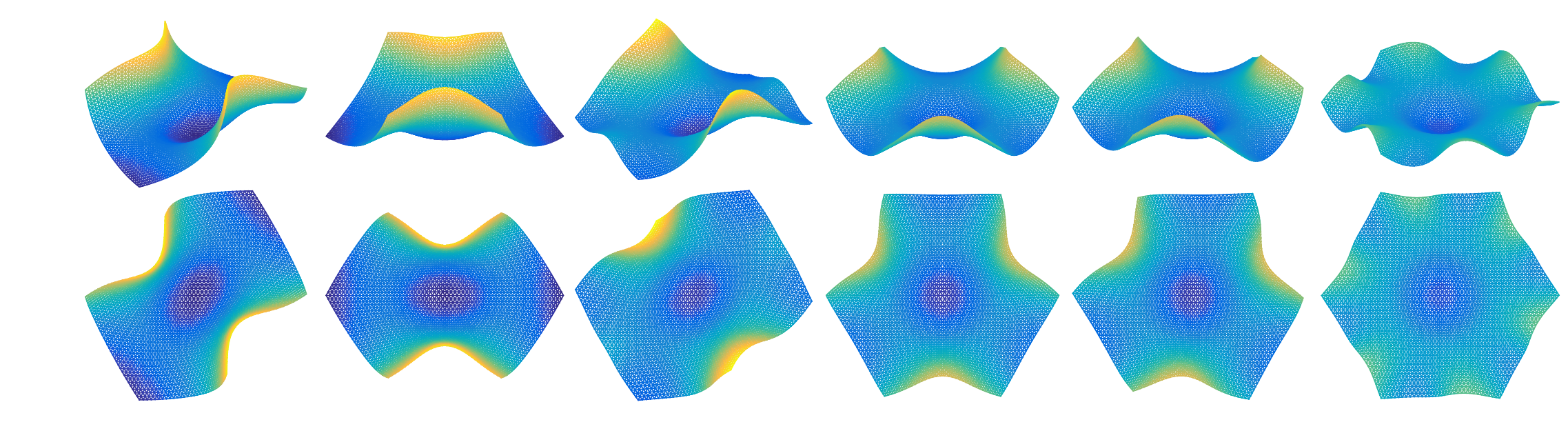} \\
           \vspace{-.25in} \hspace{-.25in}
           \end{tabular}
          \caption{\footnotesize A selection of six buckled equilibria of the unit hexagon. Each is shown in
two views: the top row gives views from 60 degrees with respect to the $z$-axis, and the
bottom row gives views along the $z$-axis. The sheet thickness
is $h = 0.03$ and the imposed reference metric factor is $\eta = \eta_1$ given in (\ref{eta1}) with
$A = 0.1$ and $k = 1$. The equilibria are obtained by starting 
from different initial perturbations, with lattice spacing $d$ = 1/33.
 \label{fig:ICFig}}
           \end{center}
         \vspace{-.10in}
        \end{figure}

Next, we study the postbuckling behavior, again with 
the static metric factor $\eta =\eta_1$ in (\ref{eta1}) but with $A = 0.1$, $k = 1$, and different initial perturbations
of the form $z = c r^m \sin(m\phi)$, where $r$ is the initial distance of sheet points from the
hexagon center, $\phi$ is the azimuthal angle (with respect to the hexagon center), and $m$ is the
azimuthal wavenumber, an integer from 2 to 6.  The amplitude $c$ ranges from 0 to 0.1, and 
$\mu$ ranges from 1 to 1000. We obtain buckling into various equilibria akin to those studied
by \cite{efrati:2009b} with other radially symmetric reference metrics. We show six equilibria
in figure \ref{fig:ICFig}, in two views (top and bottom rows). The first three (starting at the left) have
a twofold azimuthal rotational symmetry, but are distinct configurations. The second
has an additional bilateral symmetry and the third has an additional pair of local maxima
along the sheet edges. The fourth and fifth have a threefold rotational symmetry, and the fourth
has bilateral symmetries in addition. The sixth has a sixfold rotational symmetry. These are only 
a selection of local equilibria for a given static reference metric, and illustrate the complexity of
the energy landscape for such sheets, even without dynamics and/or a time-dependent reference metric.

\subsection{Parameter sweeps}

We next study the effects of parameters on the sheet dynamics, using the reference metric factor $\eta =\eta_3$ in (\ref{eta3}), a
radial traveling wave.
We perform a sequence of parameter sweeps, varying one of $A$, $h$, or $\mu$ while keeping the others
fixed. We fix the wavenumber $k$ at 1 in all cases. If $k$ is much smaller, the sheet does not buckle because the reference metric is almost uniform, and planar dilation/contraction is preferred energetically. If $k$ is much larger, buckling is also inhibited because the reference metric averaged over a local region approaches the identity tensor. As a base case, we take $A$ = 0.1, 
$\mu = 1000$, and $h$ = 0.03, and vary each physical parameter in turn with numerical parameters $d = 1/33$ and $\Delta t = 0.005$. We initialize the sheet with a small out-of-plane deflection ($z = 0.02 (x_1^4 + x_2^4)$, $-1\leq x_1 = x, x_2 = y \leq 1$).

\begin{figure} [h]
           \begin{center}
           \begin{tabular}{c}
               \includegraphics[width=6.5in]{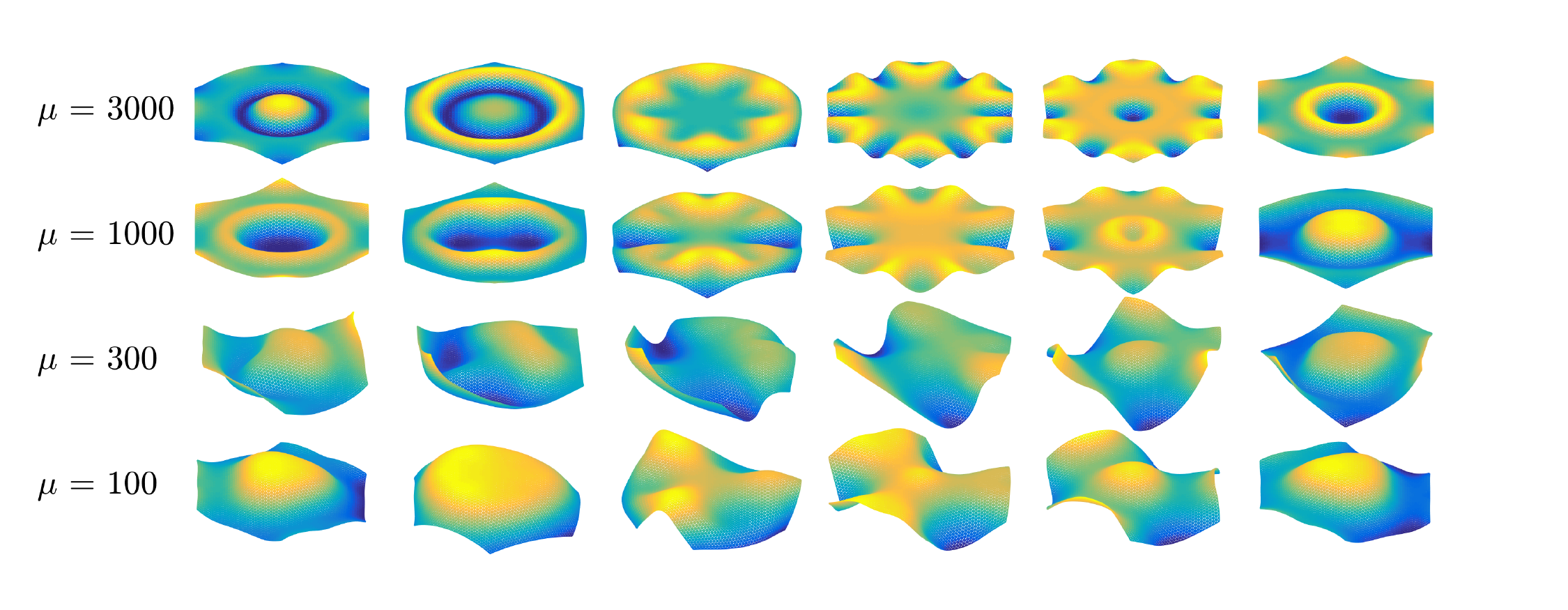} \\
           \vspace{-.25in} \hspace{-.25in}
           \end{tabular}
          \caption{\footnotesize Snapshots of a unit hexagon, from $t$ = 19 to 20 in time increments of
0.2 (from left to right) with different damping constants $\mu$ (top to bottom), 
metric factor amplitude $A$ = 0.1 in (\ref{eta3}) and $h$ = 0.03. The colors show the $z$ coordinate value, and the color
scale is scaled to the minimum and maximum $z$ coordinate value of each sheet. 
 \label{fig:muFig}}
           \end{center}
         \vspace{-.10in}
        \end{figure}

In figure \ref{fig:muFig} we show snapshots of a unit hexagon from $t$ = 19 to 20 in time increments of
0.2 (from left to right), for four
different values of $\mu$, with $A$ = 0.1 and $h$ = 0.03. When $\mu$ is large, the sheet responds more slowly to the reference metric. The tendency is to oscillate about a mean state of no deformation (because the long-time-average reference metric is the identity tensor), and at sufficiently large $\mu$ the sheet does not buckle out of plane.
At the largest $\mu$ (3000), the dynamics are essentially periodic (with period 2, so only a half-period is shown; the position at time $t+1$ is that at time $t$ but reflected in the $z = 0$ plane). The out-of-plane deflection is
smaller and more symmetric that at smaller $\mu$, where the dynamics become more chaotic and asymmetric. 
At smaller $\mu$, the sheet moves more rapidly through the complicated, time-dependent elastic energy landscape.
At $\mu = 1000$, the sheet deflection $\Delta z(t)$ has a large component with the same period as the reference metric, but also large components that are not periodic.  At $\mu = 300$, bilateral symmetry is lost, while at $\mu = 100$, the sheet almost assumes a threefold symmetry. 

\begin{figure} [h]
           \begin{center}
           \begin{tabular}{c}
               \includegraphics[width=6.5in]{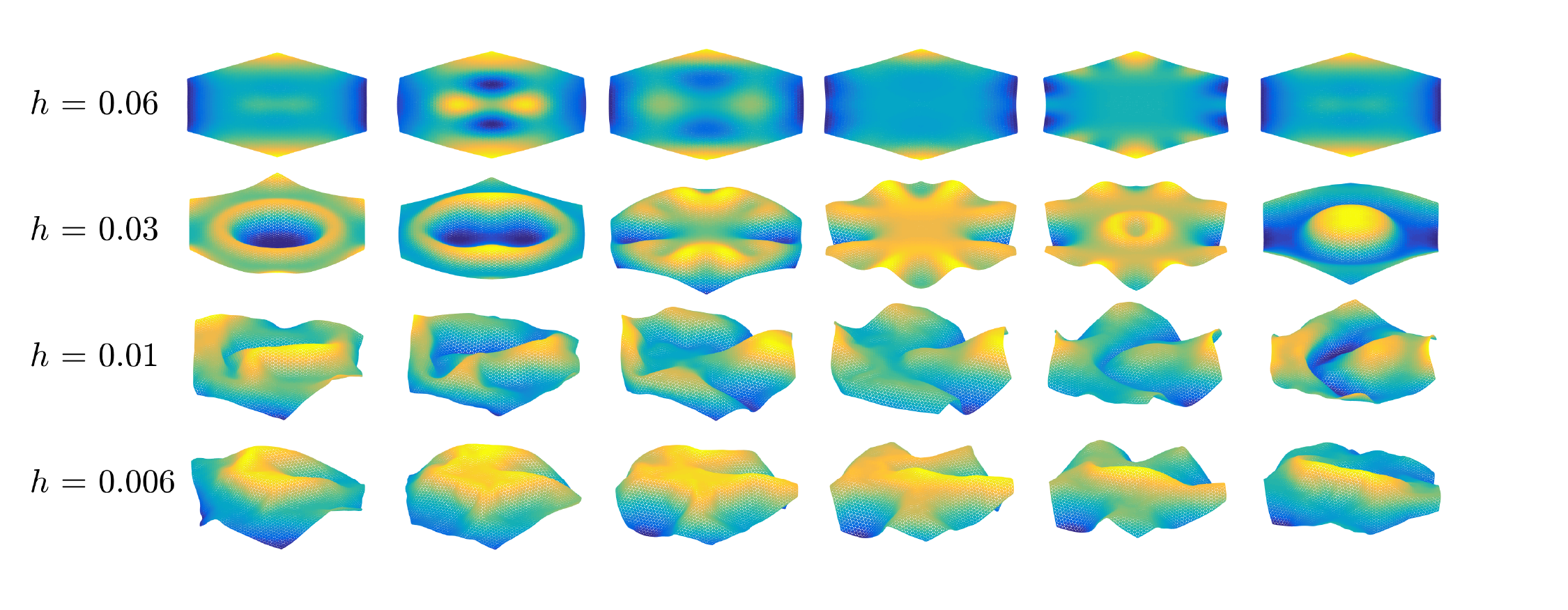} \\
           \vspace{-.25in} \hspace{-.25in}
           \end{tabular}
          \caption{\footnotesize Snapshots of sheet dynamics with various sheet thicknesses $h$ (top to bottom), from $t$ = 19 to 20 in time increments of
0.2 (from left to right). Here $A$ = 0.1 and $\mu = 1000$.
 \label{fig:hFig}}
           \end{center}
         \vspace{-.10in}
        \end{figure}

Next, we vary $h$, with $A$ = 0.1 and $\mu = 1000$. Snapshots are shown in figure  \ref{fig:hFig}.
With larger $h$, bending is relatively more costly, so the deformation is smoother, and at sufficiently large $h$ the
sheet relaxes back to a planar state ($h =0.06$). At smaller $h$, the sheet motion is more chaotic and asymmetric, as 
for decreasing $\mu$. Notably, at smaller $h$ fine wrinkling features
appear, and the deformation is far from any kind of symmetry. 
Eventually the sheet may intersect itself (not shown) because forces due to self-contact are not included.

\begin{figure} [h]
           \begin{center}
           \begin{tabular}{c}
               \includegraphics[width=6.5in]{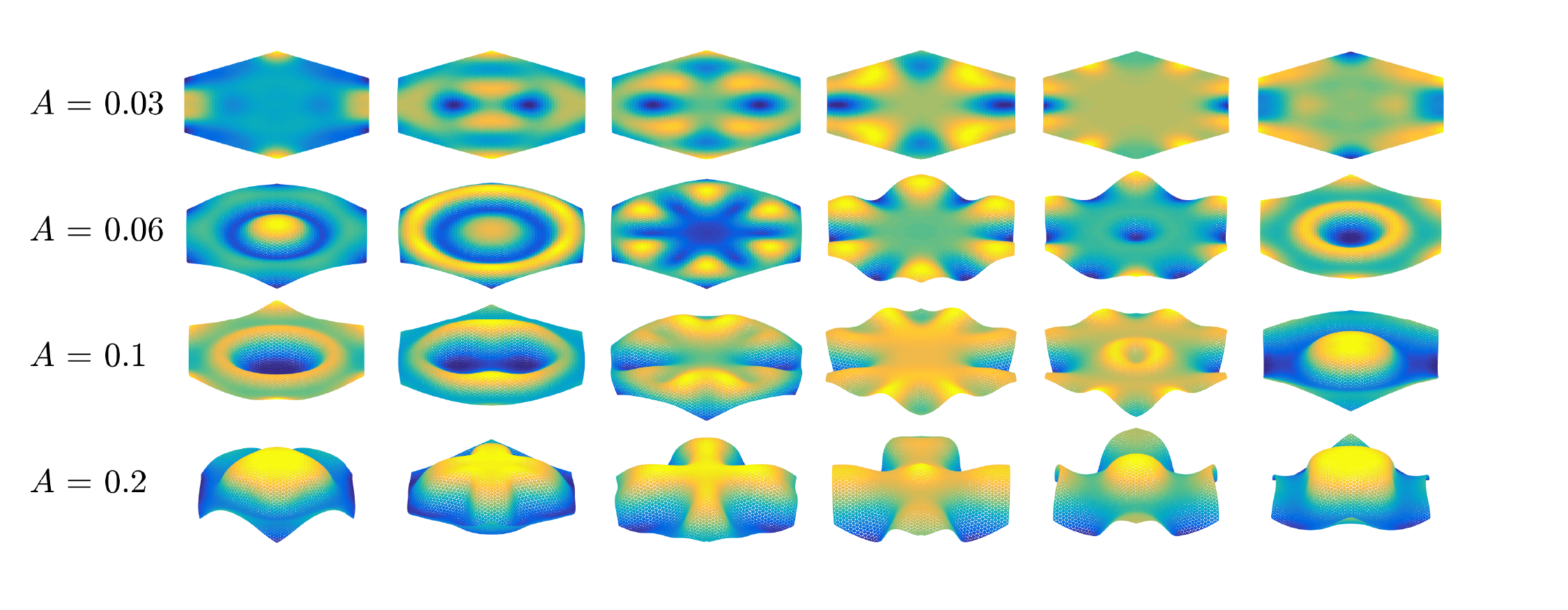} \\
           \vspace{-.25in} \hspace{-.25in}
           \end{tabular}
          \caption{\footnotesize  Snapshots of sheet dynamics with various metric factor amplitudes $A$ (in (\ref{eta3})), from $t$ = 19 to 20 in time increments of
0.2 (from left to right).
Here $h$ = 0.03 and $\mu = 1000$.
 \label{fig:AmpFig}}
           \end{center}
         \vspace{-.10in}
        \end{figure}

We vary $A$ next, with $h = 0.03$ and $\mu = 1000$. Here (figure \ref{fig:AmpFig}) 
there is a sequence of dynamics from slow relaxation to planar motions
($A$ = 0.03), to periodic and symmetric ($A$ = 0.06, with period 2), to aperiodic but still bilaterally symmetric ($A$ = 0.1 and 0.2).  
At larger $A$ the deflection amplitude increases but the deformation remains relatively smooth, and in this sense
the dynamics are more similar to smaller $\mu$ than to smaller $h$. 

\begin{figure} [h]
           \begin{center}
           \begin{tabular}{c}
               \includegraphics[width=6.5in]{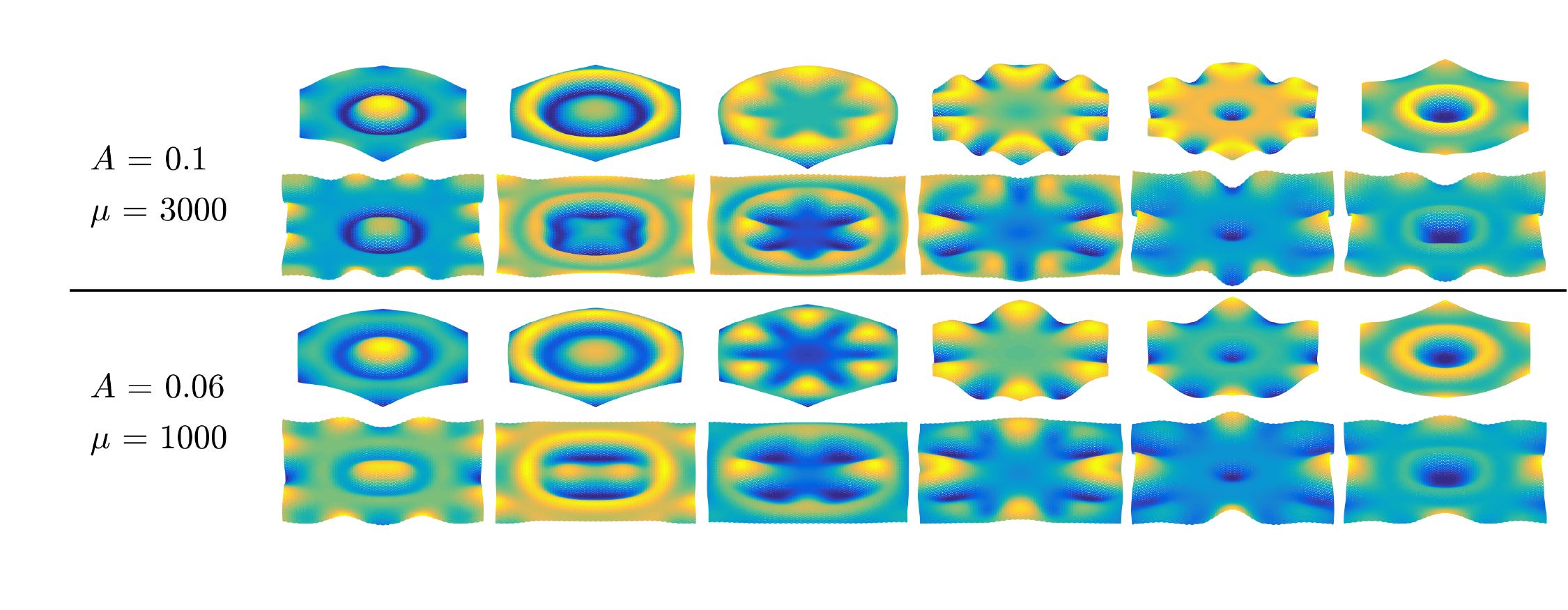} \\
           \vspace{-.25in} \hspace{-.25in}
           \end{tabular}
          \caption{\footnotesize Comparisons of the dynamics of hexagonal and square sheets with 
$A = 0.1$ and $\mu = 3000$ in the top two rows and $A = 0.06$ and $\mu = 1000$ in the bottom two rows. All
other parameters and initial conditions are the same as in figures \ref{fig:muFig} and \ref{fig:AmpFig} (e.g. $h = 0.03$),
and the snapshots run from $t$ = 19 to 20 in time increments of 0.2 (from left to right).
 \label{fig:SquareHexagonFig}}
           \end{center}
         \vspace{-.10in}
        \end{figure}

Finally, we consider the effect of the sheet shape on the dynamics. We consider the two cases with periodic dynamics above,
($A = 0.1$, $\mu = 3000$) in figure \ref{fig:muFig} and ($A = 0.06$, $\mu = 1000$) in figure \ref{fig:AmpFig}, both with $h = 0.03$. In the top two rows of figure \ref{fig:SquareHexagonFig}, we compare the hexagonal sheet snapshots with
those of a square sheet with the same parameters. Both sheets have width 2 when the reference metric is the identity. 
The bottom two rows
make the same comparison with $A = 0.06$ and $\mu = 1000$. At each time, the hexagonal and square sheets have
qualitative similarities. At the first time (leftmost snapshots), the sheets have a central hump. At the second time,
an upward ring (yellow) appears. Third, the ring breaks up into an array of humps with
sixfold symmetry for the hexagons, and either six- or fourfold symmetry for the squares. Fourth, the humps reach the sheet boundary.
Fifth, a central depression forms, and sixth, it widens and is surrounded by an upward ring. 
The main differences between the squares and
hexagons are that the squares more often have a fourfold rather than sixfold symmetry, the maximum deflections of the
squares are larger, and the squares have a significant nonperiodic component in the dynamics. The comparison illustrates
that periodic dynamics can be somewhat sensitive to sheet shape, which is perhaps not surprising given that chaotic dynamics
are common and small changes in various parameters can shift periodic dynamics to chaotic dynamics.

\section{Conclusion}

We have presented semi-implicit algorithms that can simulate the dynamics of thin elastic sheets with large time steps.
We focus on the case of elastic sheets with nontrivial steady or time-varying reference metrics in overdamped dynamics, 
but the methods should apply to other problems with internal or external forcing. 
The first algorithm simulates a triangular lattice mesh, and uses a splitting of the stretching force that has been
used previously for computer graphics simulations of hair and cloth \cite{desbrun1999interactive,selle2008mass}. The
semi-implicit bending force uses a biharmonic operator with free-edge boundary conditions. The algorithm appears
to be unconditionally stable for a periodic reference metric. The second algorithm
simulates a rectangular grid with finite-difference derivatives of the energy, and allows for general values of the Poisson
ratio (and unlike the triangular lattice, a consistent value in all terms of the bending and stretching energies). The semi-implicit finite
difference algorithm
is analogous to that for the triangular lattice, involving a stretching force operator for zero-rest-length springs, and
the biharmonic operator for an approximate bending force. The finite
difference algorithm is not unconditionally stable, but typically has a maximum stable time step two to three orders of magnitude
greater than that of an explicit scheme.

The two algorithms agree very closely for the deformation of a square sheet under a static reference metric, even allowing a
different Poisson ratio ($-1/3$ instead of 1/3) in one of the bending energy terms. There is more disagreement in the
deflection amplitude and pattern for a unidirectional traveling wave reference metric, but the dynamics
are qualitatively similar. For the triangular lattice, we see phenomena similar to those reported previously 
with different radially-symmetric reference metrics \cite{efrati:2009b}: 
a 1/2-power-law scaling of the buckling threshold in the space of sheet thickness and reference metric factor amplitude, and the presence
of multiple stable equilibria with varying types of azimuthal and bilateral symmetry. For the case of a radial
traveling wave reference metric, we showed some of the basic effects of sheet and metric parameters. In general, as the
metric factor amplitude increases, the sheet thickness decreases, or the damping parameter decreases,
the sheet moves from flat, periodic oscillations, to buckled periodic oscillations with various types of symmetries, 
to buckled motions with a combination of periodic and nonperiodic components, and varying degrees of symmetry/asymmetry.
Below a critical sheet thickness, the motions have little semblance of symmetry or periodicity.

\section*{Acknowledgments}
We acknowledge support from the 
Michigan Institute for Computational Discovery and Engineering (MICDE). 


\begin{appendix}
\section{Bending force \label{sec:BendingForce}}

We denote the discrete biharmonic operator on the triangular mesh with free-edge boundary conditions as $\tilde{\Delta}^2_{x_1,x_2}$, and because it is independent of the sheet configuration we may derive it for a sheet that is a portion of a 
flat triangular lattice with edge length $d$. 
$\tilde{\Delta}^2_{x_1,x_2}$ is a mapping from (small) out-of-plane displacements at each
point to the bending force at that point. The mapping is a sum of the force-displacement mappings for each pair of adjacent equilateral triangles, because the bending energy is also a sum over such units. For the four vertices in
a given pair of neighboring triangles, a unit upward displacement of one of the two outer vertices (those not on the shared edge)
yields a downward bending force at each of the outer vertices, equal by symmetry, and an upward force at each of the inner two vertices (those on the shared edge), equal and opposite to those at the outer 
vertices, to main net force and torque balance. The
forces are proportional to the displacement by a constant that gives the desired bending modulus of the sheet.
The bottom row of figure \ref{fig:BiharmStencilFig} shows examples of the stencils for this biharmonic operator at a few mesh points: an interior
point (analogous to the 13-point biharmonic stencil on a rectangular mesh), a next-to-boundary point,
and a boundary point. The discrete biharmonic operator $\tilde{\Delta}^2_{x_1,x_2}$ has these stencil values multiplied by the triangle altitudes raised to the $-4$ power $\left(\sqrt{3}d/2\right)^{-4}$, while the 
linearized bending force operator has instead the same
stencil values multiplied by $\left(\sqrt{3}d/2\right)^{-2}$ because the bending force is the gradient of the bending energy, which is approximately the biharmonic of the deflection multiplied by the sheet area per vertex.
\end{appendix}
\bibliographystyle{unsrt}
\bibliography{SemiImplicitSheets}
\end{document}